\documentclass[a4paper,10pt,conference]{IEEEtran}

\usepackage[
  top=19mm,
  bottom=44mm,
  left=13mm,
  right=13mm
]{geometry}

\usepackage[pdftex]{graphicx}
\usepackage{multirow}   
\usepackage{amsmath}
\usepackage{amssymb}
\usepackage{bbm}
\usepackage{eqparbox}
\usepackage{subcaption}
\usepackage{booktabs}
\usepackage{xtab}
\usepackage[hidelinks]{hyperref}
\usepackage{mathtools}
\usepackage{xcolor}
\usepackage{comment}
\usepackage{makecell}
\usepackage[acronyms,nonumberlist,nopostdot,nomain,nogroupskip,acronymlists={hidden}]{glossaries}
\usepackage[linesnumbered,ruled,vlined]{algorithm2e}
\usepackage{float} 
\usepackage[utf8]{inputenc}

\DeclareMathOperator{\diag}{diag}

\newacronym{3gpp}{3GPP}{3rd Generation Partnership Project}
\newacronym{4g}{4G}{4th generation}
\newacronym{5g}{5G}{5th generation}
\newacronym{6g}{6G}{6th generation}
\newacronym{5gc}{5GC}{5G Core}
\newacronym{adc}{ADC}{Analog to Digital Converter}
\newacronym{aerpaw}{AERPAW}{Aerial Experimentation and Research Platform for Advanced Wireless}
\newacronym{ai}{AI}{Artificial Intelligence}
\newacronym{LE}{LE}{Log-Euclidean}
\newacronym{TP}{TP}{Test Point}
\newacronym{CMD}{CMD}{Colinearity Matrix Distance}

\newacronym{EM}{EM}{electromagnetic}
\newacronym{EMS}{EMS}{electromagnetic skin}
\newacronym{mmW}{mmW}{millimeter wave}
\newacronym{RoI}{RoI}{region of interest}
\newacronym{AoA}{AoA}{Angle of Arrival}
\newacronym{AoD}{AoD}{Angle of Departure}
\newacronym{AoR}{AoR}{Angle of Reflection}
\newacronym{AoI}{AoI}{Angle of Incidence}
\newacronym{RIS}{RIS}{Reconfigurable Intelligent Surface}
\newacronym{RISs}{RISs}{Reconfigurable Intelligent Surfaces}
\newacronym{SNR}{SNR}{signal-to-noise ratio}
\newacronym{AF}{AF}{amplify-and-forward}
\newacronym{DF}{DF}{decode-and-forward}
\newacronym{STAR-RISs}{STAR-RISs}{Simultaneous Transmit and Reflecting RISs}
\newacronym{3GPP}{3GPP}{3rd Generation Partnership Project}
\newacronym{RAN}{RAN}{Radio Access Network}
\newacronym{KS}{KS}{Kruskal Stress}
\newacronym{CT}{CT}{Continuity}
\newacronym{TW}{TW}{Trustworthiness }
\newacronym{NCR}{NCR}{Network-Controlled Repeater}
\newacronym{NCRs}{NCRs}{Network-Controlled Repeaters}
\newacronym{IAB}{IAB}{Integated-Access-and-Backhauling}
\newacronym{SRE}{SRE}{Smart Radio Environment}
\newacronym{ECDF}{ECDF}{Empirical Cumulative Distribution Function}
\newacronym{CDF}{CDF}{Cumulative Distribution Function}
\newacronym{HSRE}{HSRE}{Heterogeneous SRE}
\newacronym{aimd}{AIMD}{Additive Increase Multiplicative Decrease}
\newacronym{am}{AM}{Acknowledged Mode}
\newacronym{amc}{AMC}{Adaptive Modulation and Coding}
\newacronym{FCAE}{FCAE}{Fully Connected Autoencoder}
\newacronym{amf}{AMF}{Access and Mobility Management Function}
\newacronym{aops}{AOPS}{Adaptive Order Prediction Scheduling}
\newacronym{api}{API}{Application Programming Interface}
\newacronym{apn}{APN}{Access Point Name}
\newacronym{LoS}{LoS}{Line-of-Sight}
\newacronym{NLoS}{NLoS}{None-Line-of-Sight}
\newacronym{NLDR}{NLDR}{Nonlinear Dimensionality Reduction}

\newacronym{ap}{AP}{Application Protocol}
\newacronym{ae}{ae}{Autoencoder}
\newacronym{aqm}{AQM}{Active Queue Management}
\newacronym{ar}{AR}{Augmented Reality}
\newacronym{6G}{6G}{sixth-generation}
\newacronym{MLE}{MLE}{Mean Localization Error}
\newacronym{ausf}{AUSF}{Authentication Server Function}
\newacronym{avc}{AVC}{Advanced Video Coding}
\newacronym{awgn}{AGWN}{Additive White Gaussian Noise}
\newacronym{balia}{BALIA}{Balanced Link Adaptation Algorithm}
\newacronym{bbu}{BBU}{Base Band Unit}
\newacronym{bdp}{BDP}{Bandwidth-Delay Product}
\newacronym{ber}{BER}{Bit Error Rate}
\newacronym{bf}{BF}{Beamforming}
\newacronym{bler}{BLER}{Block Error Rate}
\newacronym{OSM}{OSM}{Open Street Map}
\newacronym{MSE}{MSE}{Mean Square Error}
\newacronym{brr}{BRR}{Bayesian Ridge Regressor}
\newacronym{BS}{BS}{Base Station}
\newacronym{bsr}{BSR}{Buffer Status Report}
\newacronym{bss}{BSS}{Business Support System}
\newacronym{ca}{CA}{Carrier Aggregation}
\newacronym{caas}{CaaS}{Connectivity-as-a-Service}
\newacronym{GPS}{GPS}{Global Positioning System}
\newacronym{cb}{CB}{Code Block}
\newacronym{cc}{CC}{channel charting}
\newacronym{ccid}{CCID}{Congestion Control ID}
\newacronym{cco}{CC}{Carrier Component}
\newacronym{cdd}{CDD}{Cyclic Delay Diversity}

\newacronym{cdn}{CDN}{Content Distribution Network}
\newacronym{cn}{CN}{Core Network}
\newacronym{codel}{CoDel}{Controlled Delay Management}
\newacronym{comac}{COMAC}{Converged Multi-Access and Core}
\newacronym{cord}{CORD}{Central Office Re-architected as a Datacenter}
\newacronym{cornet}{CORNET}{COgnitive Radio NETwork}
\newacronym{cosmos}{COSMOS}{Cloud Enhanced Open Software Defined Mobile Wireless Testbed for City-Scale Deployment}
\newacronym{cots}{COTS}{Commercial Off-the-Shelf}
\newacronym{cp}{CP}{Control Plane}
\newacronym{cyp}{CP}{Cyclic Prefix}
\newacronym{up}{UP}{User Plane}
\newacronym{cpu}{CPU}{Central Processing Unit}
\newacronym{cqi}{CQI}{Channel Quality Information}
\newacronym{cr}{CR}{Cognitive Radio}
\newacronym{cran}{C-RAN}{Cloud \gls{ran}}
\newacronym{crs}{CRS}{Cell Reference Signal}
\newacronym{csi}{CSI}{Channel State Information}
\newacronym{csirs}{CSI-RS}{Channel State Information - Reference Signal}
\newacronym{cu}{CU}{Central Unit}
\newacronym{d2tcp}{D$^2$TCP}{Deadline-aware Data center TCP}
\newacronym{d3}{D$^3$}{Deadline-Driven Delivery}
\newacronym{dac}{DAC}{Digital to Analog Converter}
\newacronym{dag}{DAG}{Directed Acyclic Graph}
\newacronym{das}{DAS}{Distributed Antenna System}
\newacronym{dash}{DASH}{Dynamic Adaptive Streaming over HTTP}
\newacronym{dc}{DC}{Dual Connectivity}
\newacronym{dccp}{DCCP}{Datagram Congestion Control Protocol}
\newacronym{dce}{DCE}{Direct Code Execution}
\newacronym{dci}{DCI}{Downlink Control Information}
\newacronym{dctcp}{DCTCP}{Data Center TCP}
\newacronym{dl}{DL}{Downlink}
\newacronym{dmr}{DMR}{Deadline Miss Ratio}
\newacronym{dmrs}{DMRS}{DeModulation Reference Signal}
\newacronym{drlcc}{DRL-CC}{Deep Reinforcement Learning Congestion Control}
\newacronym{drs}{DRS}{Discovery Reference Signal}
\newacronym{du}{DU}{Distributed Unit}
\newacronym{e2e}{E2E}{end-to-end}
\newacronym{ecaas}{ECaaS}{Edge-Cloud-as-a-Service}
\newacronym{ecn}{ECN}{Explicit Congestion Notification}
\newacronym{edf}{EDF}{Earliest Deadline First}
\newacronym{embb}{eMBB}{Enhanced Mobile Broadband}
\newacronym{empower}{EMPOWER}{EMpowering transatlantic PlatfOrms for advanced WirEless Research}
\newacronym{enb}{eNB}{evolved Node Base}
\newacronym{endc}{EN-DC}{E-UTRAN-\gls{nr} \gls{dc}}
\newacronym{epc}{EPC}{Evolved Packet Core}
\newacronym{eps}{EPS}{Evolved Packet System}
\newacronym{es}{ES}{Edge Server}
\newacronym{etsi}{ETSI}{European Telecommunications Standards Institute}
\newacronym[firstplural=Estimated Times of Arrival (ETAs)]{eta}{ETA}{Estimated Time of Arrival}
\newacronym{eutran}{E-UTRAN}{Evolved Universal Terrestrial Access Network}
\newacronym{faas}{FaaS}{Function-as-a-Service}
\newacronym{fapi}{FAPI}{Functional Application Platform Interface}
\newacronym{fdd}{FDD}{Frequency Division Duplexing}
\newacronym{fdm}{FDM}{Frequency Division Multiplexing}
\newacronym{fdma}{FDMA}{Frequency Division Multiple Access}
\newacronym{fed4fire}{FED4FIRE+}{Federation 4 Future Internet Research and Experimentation Plus}
\newacronym{fir}{FIR}{Finite Impulse Response}
\newacronym{fit}{FIT}{Future \acrlong{iot}}
\newacronym{fpga}{FPGA}{Field Programmable Gate Array}
\newacronym{fr2}{FR2}{Frequency Range 2}
\newacronym{fs}{FS}{Fast Switching}
\newacronym{fscc}{FSCC}{Flow Sharing Congestion Control}
\newacronym{ftp}{FTP}{File Transfer Protocol}
\newacronym{fw}{FW}{Flow Window}
\newacronym{ge}{GE}{Gaussian Elimination}
\newacronym{gnb}{gNB}{Next Generation Node Base}
\newacronym{gop}{GOP}{Group of Pictures}
\newacronym{gpr}{GPR}{Gaussian Process Regressor}
\newacronym{gpu}{GPU}{Graphics Processing Unit}
\newacronym{gtp}{GTP}{GPRS Tunneling Protocol}
\newacronym{gtpc}{GTP-C}{GPRS Tunnelling Protocol Control Plane}
\newacronym{gtpu}{GTP-U}{GPRS Tunnelling Protocol User Plane}
\newacronym{gtpv2c}{GTPv2-C}{\gls{gtp} v2 - Control}
\newacronym{gw}{GW}{Gateway}
\newacronym{harq}{HARQ}{Hybrid Automatic Repeat reQuest}
\newacronym{hetnet}{HetNet}{Heterogeneous Network}
\newacronym{hh}{HH}{Hard Handover}
\newacronym{hol}{HOL}{Head-of-Line}
\newacronym{hqf}{HQF}{Highest-quality-first}
\newacronym{hss}{HSS}{Home Subscription Server}
\newacronym{http}{HTTP}{HyperText Transfer Protocol}
\newacronym{ia}{IA}{Initial Access}
\newacronym{iab}{IAB}{Integrated Access and Backhaul}
\newacronym{ic}{IC}{Incident Command}
\newacronym{ietf}{IETF}{Internet Engineering Task Force}
\newacronym{imsi}{IMSI}{International Mobile Subscriber Identity}
\newacronym{imt}{IMT}{International Mobile Telecommunication}
\newacronym{iot}{IoT}{Internet of Things}
\newacronym{ip}{IP}{Internet Protocol}
\newacronym{itu}{ITU}{International Telecommunication Union}
\newacronym{kpi}{KPI}{Key Performance Indicator}
\newacronym{kpm}{KPM}{Key Performance Measurement}
\newacronym{kvm}{KVM}{Kernel-based Virtual Machine}
\newacronym{los}{LOS}{Line-of-Sight}
\newacronym{lsm}{LSM}{Link-to-System Mapping}
\newacronym{lstm}{LSTM}{Long Short Term Memory}
\newacronym{lte}{LTE}{Long Term Evolution}
\newacronym{lxc}{LXC}{Linux Container}
\newacronym{m2m}{M2M}{Machine to Machine}
\newacronym{mac}{MAC}{Medium Access Control}
\newacronym{manet}{MANET}{Mobile Ad Hoc Network}
\newacronym{mano}{MANO}{Management and Orchestration}
\newacronym{mc}{MC}{Multi-Connectivity}
\newacronym{mcc}{MCC}{Mobile Cloud Computing}
\newacronym{mchem}{MCHEM}{Massive Channel Emulator}
\newacronym{mcs}{MCS}{Modulation and Coding Scheme}
\newacronym{mec}{MEC}{Multi-access Edge Computing}
\newacronym{mec2}{MEC}{Mobile Edge Cloud}
\newacronym{mfc}{MFC}{Mobile Fog Computing}
\newacronym{mgen}{MGEN}{Multi-Generator}
\newacronym{mi}{MI}{Mutual Information}
\newacronym{mib}{MIB}{Master Information Block}
\newacronym{miesm}{MIESM}{Mutual Information Based Effective SINR}
\newacronym{mimo}{MIMO}{Multiple Input, Multiple Output}
\newacronym{ml}{ML}{Machine Learning}
\newacronym{mlr}{MLR}{Maximum-local-rate}
\newacronym[plural=\gls{mme}s,firstplural=Mobility Management Entities (MMEs)]{mme}{MME}{Mobility Management Entity}
\newacronym{mmtc}{mMTC}{Massive Machine-Type Communications}
\newacronym{mmwave}{mmWave}{millimeter wave}
\newacronym{mpdccp}{MP-DCCP}{Multipath Datagram Congestion Control Protocol}
\newacronym{mptcp}{MPTCP}{Multipath TCP}
\newacronym{mr}{MR}{Maximum Rate}
\newacronym{mrdc}{MR-DC}{Multi \gls{rat} \gls{dc}}
\newacronym{mse}{MSE}{Mean Square Error}
\newacronym{mss}{MSS}{Maximum Segment Size}
\newacronym{mt}{MT}{Mobile Termination}
\newacronym{mtd}{MTD}{Machine-Type Device}
\newacronym{mtu}{MTU}{Maximum Transmission Unit}
\newacronym{mumimo}{MU-MIMO}{Multi-user \gls{mimo}}
\newacronym{mvno}{MVNO}{Mobile Virtual Network Operator}
\newacronym{nalu}{NALU}{Network Abstraction Layer Unit}
\newacronym{nas}{NAS}{Non-Access Stratum}
\newacronym{nbiot}{NB-IoT}{Narrow Band IoT}
\newacronym{nfv}{NFV}{Network Function Virtualization}
\newacronym{nfvi}{NFVI}{Network Function Virtualization Infrastructure}
\newacronym{ngrg}{nGRG}{next Generation Research Group}
\newacronym{ni}{NI}{Network Interfaces}
\newacronym{nic}{NIC}{Network Interface Card}
\newacronym{nlos}{NLOS}{Non-Line-of-Sight}
\newacronym{now}{NOW}{Non Overlapping Window}
\newacronym{nsm}{NSM}{Network Service Mesh}
\newacronym{nr}{NR}{New Radio}
\newacronym{nrf}{NRF}{Network Repository Function}
\newacronym{nsa}{NSA}{Non Stand Alone}
\newacronym{nse}{NSE}{Network Slicing Engine}
\newacronym{nssf}{NSSF}{Network Slice Selection Function}
\newacronym{o2i}{O2I}{Outdoor to Indoor}
\newacronym{oai}{OAI}{OpenAirInterface}
\newacronym{oaicn}{OAI-CN}{\gls{oai} \acrlong{cn}}
\newacronym{oairan}{OAI-RAN}{\acrlong{oai} \acrlong{ran}}
\newacronym{oam}{OAM}{Operations, Administration and Maintenance}
\newacronym{ofdm}{OFDM}{Orthogonal Frequency Division Multiplexing}
\newacronym{olia}{OLIA}{Opportunistic Linked Increase Algorithm}
\newacronym{omec}{OMEC}{Open Mobile Evolved Core}
\newacronym{onap}{ONAP}{Open Network Automation Platform}
\newacronym{onf}{ONF}{Open Networking Foundation}
\newacronym{onos}{ONOS}{Open Networking Operating System}
\newacronym{oom}{OOM}{\gls{onap} Operations Manager}
\newacronym{opnfv}{OPNFV}{Open Platform for \gls{nfv}}
\newacronym{oran}{O-RAN}{Open Radio Access Network}
\newacronym{orbit}{ORBIT}{Open-Access Research Testbed for Next-Generation Wireless Networks}
\newacronym{os}{OS}{Operating System}
\newacronym{oss}{OSS}{Operations Support System}
\newacronym{otic}{OTIC}{Open Testing \& Integration Centre}
\newacronym{pa}{PA}{Position-aware}
\newacronym{pase}{PASE}{Prioritization, Arbitration, and Self-adjusting Endpoints}
\newacronym{pawr}{PAWR}{Platforms for Advanced Wireless Research}
\newacronym{pbch}{PBCH}{Physical Broadcast Channel}
\newacronym{pcef}{PCEF}{Policy and Charging Enforcement Function}
\newacronym{pcfich}{PCFICH}{Physical Control Format Indicator Channel}
\newacronym{pcrf}{PCRF}{Policy and Charging Rules Function}
\newacronym{pdcch}{PDCCH}{Physical Downlink Control Channel}
\newacronym{pdcp}{PDCP}{Packet Data Convergence Protocol}
\newacronym{pdsch}{PDSCH}{Physical Downlink Shared Channel}
\newacronym{pdu}{PDU}{Packet Data Unit}
\newacronym{pf}{PF}{Proportional Fair}
\newacronym{pgw}{PGW}{Packet Gateway}
\newacronym{phich}{PHICH}{Physical Hybrid ARQ Indicator Channel}
\newacronym{phy}{PHY}{Physical}
\newacronym{pmch}{PMCH}{Physical Multicast Channel}
\newacronym{pmi}{PMI}{Precoding Matrix Indicators}
\newacronym{powder}{POWDER}{Platform for Open Wireless Data-driven Experimental Research}
\newacronym{ppo}{PPO}{Proximal Policy Optimization}
\newacronym{ppp}{PPP}{Poisson Point Process}
\newacronym{prach}{PRACH}{Physical Random Access Channel}
\newacronym{prb}{PRB}{Physical Resource Block}
\newacronym{psnr}{PSNR}{Peak Signal to Noise Ratio}
\newacronym{pss}{PSS}{Primary Synchronization Signal}
\newacronym{pucch}{PUCCH}{Physical Uplink Control Channel}
\newacronym{pusch}{PUSCH}{Physical Uplink Shared Channel}
\newacronym{rar}{RAR}{Random Access Response}
\newacronym{qam}{QAM}{Quadrature Amplitude Modulation}
\newacronym{qci}{QCI}{\gls{qos} Class Identifier}
\newacronym{5qi}{5QI}{5G \gls{qos} Identifier}
\newacronym{qoe}{QoE}{Quality of Experience}
\newacronym{QoS}{QoS}{Quality of Service}
\newacronym{UE}{UE}{User Equipment}
\newacronym{UEs}{UEs}{User Equipments}
\newacronym{FoV}{FoV}{field of view}
\newacronym{UPA}{UPA}{uniform planar array}
\newacronym{quic}{QUIC}{Quick UDP Internet Connections}
\newacronym{rach}{RACH}{Random Access Channel}
\newacronym{ran}{RAN}{Radio Access Network}
\newacronym[firstplural=Radio Access Technologies (RATs)]{rat}{RAT}{Radio Access Technology}
\newacronym{rcn}{RCN}{Research Coordination Network}
\newacronym{STAR}{STAR-RIS}{simultaneous transmitting and reflecting RIS}
\newacronym{3SNCR}{3SNCR}{trisectoral NCR}
\newacronym{rc}{RC}{RAN Control}
\newacronym{rec}{REC}{Radio Edge Cloud}
\newacronym{red}{RED}{Random Early Detection}
\newacronym{renew}{RENEW}{Reconfigurable Eco-system for Next-generation End-to-end Wireless}
\newacronym{rf}{RF}{Radio Frequency}
\newacronym{rfc}{RFC}{Request for Comments}
\newacronym{rfr}{RFR}{Random Forest Regressor}
\newacronym{ric}{RIC}{\gls{ran} Intelligent Controller}
\newacronym{rlc}{RLC}{Radio Link Control}
\newacronym{rlf}{RLF}{Radio Link Failure}
\newacronym{rlnc}{RLNC}{Random Linear Network Coding}
\newacronym{rmr}{RMR}{RIC Message Router}
\newacronym{rmse}{RMSE}{Root Mean Squared Error}
\newacronym{rnis}{RNIS}{Radio Network Information Service}
\newacronym{rr}{RR}{Round Robin}
\newacronym{rrc}{RRC}{Radio Resource Control}
\newacronym{rrm}{RRM}{Radio Resource Management}
\newacronym{rru}{RRU}{Remote Radio Unit}
\newacronym{rs}{RS}{Remote Server}
\newacronym{rsrp}{RSRP}{Reference Signal Received Power}
\newacronym{rsrq}{RSRQ}{Reference Signal Received Quality}
\newacronym{rss}{RSS}{Received Signal Strength}
\newacronym{rssi}{RSSI}{Received Signal Strength Indicator}
\newacronym{rtt}{RTT}{Round Trip Time}
\newacronym{ru}{RU}{Radio Unit}
\newacronym{rw}{RW}{Receive Window}
\newacronym{rx}{RX}{Receiver}
\newacronym{s1ap}{S1AP}{S1 Application Protocol}
\newacronym{sa}{SA}{standalone}
\newacronym{sack}{SACK}{Selective Acknowledgment}
\newacronym{sap}{SAP}{Service Access Point}
\newacronym{sc2}{SC2}{Spectrum Collaboration Challenge}
\newacronym{scef}{SCEF}{Service Capability Exposure Function}
\newacronym{sch}{SCH}{Secondary Cell Handover}
\newacronym{scoot}{SCOOT}{Split Cycle Offset Optimization Technique}
\newacronym{sctp}{SCTP}{Stream Control Transmission Protocol}
\newacronym{sdap}{SDAP}{Service Data Adaptation Protocol}
\newacronym{sdk}{SDK}{Software Development Kit}
\newacronym{sdm}{SDM}{Space Division Multiplexing}
\newacronym{sdma}{SDMA}{Spatial Division Multiple Access}
\newacronym{sdn}{SDN}{Software-defined Networking}
\newacronym{sdr}{SDR}{Software-defined Radio}
\newacronym{seba}{SEBA}{SDN-Enabled Broadband Access}
\newacronym{sgsn}{SGSN}{Serving GPRS Support Node}
\newacronym{sgw}{SGW}{Service Gateway}
\newacronym{si}{SI}{Study Item}
\newacronym{sib}{SIB}{Secondary Information Block}
\newacronym{sinr}{SINR}{Signal to Interference plus Noise Ratio}
\newacronym{sip}{SIP}{Session Initiation Protocol}
\newacronym{siso}{SISO}{Single Input, Single Output}
\newacronym{sla}{SLA}{Service Level Agreement}
\newacronym{sm}{SM}{Service Model}
\newacronym{smf}{SMF}{Session Management Function}
\newacronym{smo}{SMO}{Service Management and Orchestration}
\newacronym{sms}{SMS}{Short Message Service}
\newacronym{smsgmsc}{SMS-GMSC}{\gls{sms}-Gateway}
\newacronym{snr}{SNR}{Signal-to-Noise-Ratio}
\newacronym{son}{SON}{Self-Organizing Network}
\newacronym{sptcp}{SPTCP}{Single Path TCP}
\newacronym{srb}{SRB}{Service Radio Bearer}
\newacronym{srn}{SRN}{Standard Radio Node}
\newacronym{srs}{SRS}{Sounding Reference Signal}
\newacronym{zc}{ZC}{Zadoff-Chu}
\newacronym{ta}{TA}{Timing Advance}
\newacronym{ss}{SS}{Synchronization Signal}
\newacronym{sss}{SSS}{Secondary Synchronization Signal}
\newacronym{st}{ST}{Spanning Tree}
\newacronym{svc}{SVC}{Scalable Video Coding}
\newacronym{tb}{TB}{Transport Block}
\newacronym{tcp}{TCP}{Transmission Control Protocol}
\newacronym{tdd}{TDD}{Time Division Duplexing}
\newacronym{tdm}{TDM}{Time Division Multiplexing}
\newacronym{tdma}{TDMA}{Time Division Multiple Access}
\newacronym{tfl}{TfL}{Transport for London}
\newacronym{tfrc}{TFRC}{TCP-Friendly Rate Control}
\newacronym{tft}{TFT}{Traffic Flow Template}
\newacronym{tgen}{TGEN}{Traffic Generator}
\newacronym{tip}{TIP}{Telecom Infra Project}
\newacronym{tm}{TM}{Transparent Mode}
\newacronym{to}{TO}{Telco Operator}
\newacronym{tr}{TR}{Technical Report}
\newacronym{trp}{TRP}{Transmitter Receiver Pair}
\newacronym{ts}{TS}{Technical Specification}
\newacronym{tti}{TTI}{Transmission Time Interval}
\newacronym{ttt}{TTT}{Time-to-Trigger}
\newacronym{tx}{TX}{Transmitter}
\newacronym{uas}{UAS}{Unmanned Aerial System}
\newacronym{uav}{UAV}{Unmanned Aerial Vehicle}
\newacronym{udm}{UDM}{Unified Data Management}
\newacronym{udp}{UDP}{User Datagram Protocol}
\newacronym{udr}{UDR}{Unified Data Repository}
\newacronym{ue}{UE}{User Equipment}
\newacronym{uhd}{UHD}{\gls{usrp} Hardware Driver}
\newacronym{ul}{UL}{Uplink}
\newacronym{um}{UM}{Unacknowledged Mode}
\newacronym{uml}{UML}{Unified Modeling Language}
\newacronym{upa}{UPA}{Uniform Planar Array}
\newacronym{upf}{UPF}{User Plane Function}
\newacronym{urllc}{URLLC}{Ultra Reliable and Low Latency Communications}
\newacronym{usa}{U.S.}{United States}
\newacronym{usim}{USIM}{Universal Subscriber Identity Module}
\newacronym{usrp}{USRP}{Universal Software Radio Peripheral}
\newacronym{utc}{UTC}{Urban Traffic Control}
\newacronym{vim}{VIM}{Virtualization Infrastructure Manager}
\newacronym{vm}{VM}{Virtual Machine}
\newacronym{vnf}{VNF}{Virtual Network Function}
\newacronym{volte}{VoLTE}{Voice over \gls{lte}}
\newacronym{voltha}{VOLTHA}{Virtual OLT HArdware Abstraction}
\newacronym{vr}{VR}{Virtual Reality}
\newacronym{vran}{vRAN}{Virtualized \gls{ran}}
\newacronym{vss}{VSS}{Video Streaming Server}
\newacronym{wbf}{WBF}{Wired Bias Function}
\newacronym{wf}{WF}{Waterfilling}
\newacronym{wg}{WG}{Working Group}
\newacronym{wlan}{WLAN}{Wireless Local Area Network}
\newacronym{osm}{OSM}{Open Source Management and Orchestration}
\newacronym{pnf}{PNF}{Physical Network Function}
\newacronym{drl}{DRL}{Deep Reinforcement Learning}
\newacronym{mtc}{MTC}{Machine-type Communications}
\newacronym{osc}{OSC}{O-RAN Software Community}
\newacronym{mns}{MnS}{Management Services}
\newacronym{ves}{VES}{\gls{vnf} Event Stream}
\newacronym{ei}{EI}{Enrichment Information}
\newacronym{fh}{FH}{Fronthaul}
\newacronym{fft}{FFT}{Fast Fourier Transform}
\newacronym{laa}{LAA}{Licensed-Assisted Access}
\newacronym{plfs}{PLFS}{Physical Layer Frequency Signals}
\newacronym{ptp}{PTP}{Precision Time Protocol}
\newacronym{asic}{ASIC}{Application-specific Integrated Circuit}
\newacronym{aal}{AAL}{Acceleration Abstraction Layer}
\newacronym{fec}{FEC}{Forward Error Correction}
\newacronym{sdl}{SDL}{Shared Data Layer}
\newacronym{nib}{NIB}{Network Information Base}
\newacronym{rnib}{R-NIB}{RAN \gls{nib}}
\newacronym{fcaps}{FCAPS}{Fault, Configuration, Accounting, Performance, Security}
\newacronym{ie}{IE}{Information Element}
\newacronym{fg}{FG}{Focus Group}
\newacronym{osfg}{OSFG}{Open Source Focus Group}
\newacronym{sdfg}{SDFG}{Standard Development Focus Group}
\newacronym{tifg}{TIFG}{Test \& Integration Focus Group}
\newacronym{sfg}{SFG}{Security Focus Group}
\newacronym{swg}{SWG}{Security Work Group}
\newacronym{e2sm}{E2SM}{E2 Service Model}
\newacronym{tsc}{TSC}{Technical Steering Committee}
\newacronym{sdo}{SDO}{Standard-Development Organization}
\newacronym{sql}{SQL}{Structured Query Language}
\newacronym{ssh}{SSH}{Secure Shell}
\newacronym{tls}{TLS}{Transport Layer Security}
\newacronym{netconf}{NETCONF}{Network Configuration Protocol}
\newacronym{dtls}{DTLS}{Datagram Transport Layer Security}
\newacronym{cmp}{CMP}{Certificate Management Protocol}
\newacronym{ccc}{CCC}{Cell Configuration and Control}
\newacronym{dsp}{DSP}{Digital Signal Processing}
\newacronym{opex}{OPEX}{Operational Expenses}
\newacronym{cbrs}{CBRS}{Citizen Broadband Radio Service}
\newacronym{ntn}{NTN}{Non-terrestrial Network}
\newacronym{gbr}{GBR}{Guaranteed Bitrate}
\newacronym{sps}{SPS}{Semi-Persistent Scheduling}
\newacronym{tbs}{TBS}{Transport Block Size}
\newacronym{gnss}{GNSS}{Global Navigation Satellite System}
\newacronym{tof}{ToF}{Time of Flight}
\newacronym{rtof}{RToF}{Return Time of Flight}
\newacronym{rsig}{RS}{Reference Signal}
\newacronym{nrtric}{near-RT RIC}{near-Real Time Ran Intelligent Controller}
\newacronym{nonrtric}{non-RT RIC}{non-Real Time Ran Intelligent Controller}
\newacronym{aoa}{AoA}{Angle of Arrival}
\newacronym{tdoa}{TDoA}{Time Difference of Arrival}
\newacronym{rtoa}{RToA}{Return Time of Arrival}
\newacronym{ris}{RIS}{Reconfigurable Intelligent Surface}
\newacronym{srd}{SRD}{Smart Radio Device}
\newacronym{gfbr}{GFBR}{Guaranteed Flow Bit Rate}
\newacronym{rg}{RG}{Resource Grid}
\newacronym{rb}{RB}{Resource Block}
\newacronym{re}{RE}{Resource Element}
\newacronym{rfra}{RF}{Radio Frame}
\newacronym{scs}{SCS}{Subcarrier Spacing}
\newacronym{ec}{EC}{Edge Computing}
\newacronym{af}{AF}{Amplify-and-Forward}
\newacronym{ncr}{NCR}{Network-Controlled Repeater}
\newacronym{tp}{TP}{Test Point}
\newacronym{cs}{CS}{Candidate Site}
\newacronym{src}{SRC}{Smart Radio Connection}
\newacronym{milp}{MILP}{Mixed Integer-Linear Programming}

\newacronym{FCMC}{FCMC}{full coverage minimum cost}

\newacronym{MBCC}{MBCC}{maximum budget-constrained coverage}

\newacronym{PDF}{PDF}{probability density function}
\newacronym{tsne}{t-SNE}{t-distributed stochastic neighbor 
embedding}
\newacronym{stsne}{St-SNE}{Semi-Supervised t-distributed stochastic neighbor 
embedding}

\begin{document}

\title{Channel Charting in Smart Radio Environments}

\author{Mahdi~Maleki, Reza~Agahzadeh~Ayoubi,~\IEEEmembership{Member,~IEEE,}
        Marouan~Mizmizi,~\IEEEmembership{Member,~IEEE,}
        Umberto~Spagnolini,~\IEEEmembership{Senior~Member,~IEEE}
\thanks{The authors are with the Department of Electronics, Information and Bioengineering, Politecnico di Milano, 20133, Milano, Italy}}

\maketitle

\begin{abstract}
This paper introduces the use of static \glspl{EMS} to enable robust device localization via \gls{cc} in realistic urban environments. We develop a rigorous optimization framework that leverages EMS to enhance channel dissimilarity and spatial fingerprinting, formulating EMS phase profile design as a codebook-based problem targeting the upper quantiles of key embedding metrics—localization error, trustworthiness, and continuity. Through 3D ray-traced simulations of a representative city scenario, we demonstrate that optimized \gls{EMS} configurations, in addition to significant improvement of the average positioning error, reduce the 90th-percentile localization error from over 60\,m (no \gls{EMS}) to less than 25\,m, while drastically improving trustworthiness and continuity. To the best of our knowledge, this is the first work to exploit \gls{SRE} with static EMS for enhancing \gls{cc}, achieving substantial gains in localization performance under challenging \gls{NLoS} conditions.
\end{abstract}
\glsresetall


\section{Introduction}
Wireless \gls{cc} represents a transformative approach to understanding and utilizing the intrinsic properties of wireless communication environments \cite{ferrand2023wireless}. By creating a low-dimensional representation (the chart) of high-dimensional \gls{csi}, \gls{cc} provides a framework to organize and interpret wireless channels' complex spatial and temporal characteristics. This capability unlocks diverse applications such as device localization, \cite{mmWLocalization} network optimization and etc. Conventional localization methods, such as GPS or radio-access-based techniques, face notable challenges. Their accuracy is degraded in dense urban areas due to multipath propagation and \gls{NLoS} conditions, which distort signals and introduce delays~\cite{gupta2015survey}. Moreover, precise positioning typically demands tight synchronization, calibrated hardware, and dedicated signaling, adding complexity and cost~\cite{mmWLocalization}. \gls{cc} addresses these limitations by leveraging the spatial information embedded in \gls{csi}. It benefits from multipath propagation, tolerates hardware impairments, and operates with coarse synchronization based on timing advance~\cite{ferrand2023wireless}.

\gls{cc} builds low-dimensional representations of high-dimensional \gls{csi} using non-linear dimensionality reduction methods that preserve geometric structure. Techniques like Isomap~\cite{ISOmap}, t-distributed stochastic neighbor embedding (t-SNE)~\cite{kazemi2023beam}, and autoencoders~\cite{studer2018channel} are commonly employed. Among these, t-SNE stands out for capturing subtle spatial relationships with a good trade-off between interpretability and performance.

Despite their strengths, these techniques map CSI to a latent space rather than to physical coordinates. Supervised learning can bridge this gap if position labels are available, but large-scale labeling is often impractical due to cost, privacy, and logistics~\cite{taner2025channel}. Semi-supervised approaches~\cite{zhang2021semi} mitigate this by using a small number of labeled UEs (e.g., GPS-equipped devices) to generalize localization to others.

High-frequency bands, especially \gls{mmW}, are promising for 6G due to their high bandwidth~\cite{kazemi2023beam}. However, their sparse, low-rank channels---caused by attenuation and blockage---limit spatial diversity, challenging \gls{cc}’s ability to distinguish nearby user locations~\cite{shahmansoori2017position}.

To address this, \gls{SRE} aim to shape wireless propagation using engineered metasurfaces. The most well-known type of these surfaces are known as \gls{RIS}, normally utilized for  coverage and blockage mitigation~\cite{RIS_vs_NCR}. \gls{RIS}  entail high cost and complexity due to their active reconfiguration infrastructure~\cite{ayanoglu2022wave}. More importantly, these reconfigurable surfaces necessitate prior knowledge of user positions to function, which stands in direct contrast to the goal of localization. This circular dependency—a classic chicken-and-egg dilemma—makes \glspl{RIS} incompatible with the principles of channel charting.\footnote{Note that there are works such as~\cite{Bennis_CC_for_RIS} where \gls{cc} is employed to assist \glspl{RIS} configuration, but not the reverse—i.e., metasurfaces enhancing \gls{cc}.}

Alternatively, \glspl{EMS} offer a low-cost, passive solution. These static metasurfaces reflect signals according to generalized Snell’s law~\cite{SmartSkin}, requiring no reconfiguration post-deployment and costing only a few dollars per square meter~\cite{smsk_electronics}.

To the best of our knowledge, this is the first work to apply SREs with EMS to CC. By enriching the multipath environment, we enhance the spatial dissimilarity of CSI and improve localization performance using CC.

This work makes four key contributions. First, we introduce a scenario-driven framework for optimizing static \gls{EMS} phase profiles to enhance \gls{cc}-based localization in realistic urban settings. Second, we formulate EMS design as a codebook-based optimization problem, directly targeting embedding metrics such as localization error, trustworthiness, and continuity. Third, we develop a quantile-based evaluation protocol that emphasizes worst-case user performance, especially under NLoS conditions. Lastly, we demonstrate the framework via 3D ray-traced simulations of a real urban map, comparing t-SNE-based \gls{cc} performance with and without EMS deployment.

\section{System Model}
\label{sec:system_model}

Consider the wireless communication scenario illustrated in Fig.~\ref{fig:sysmodel}, which comprises a \gls{BS} located at position $\mathbf{P}_{\mathrm{BS}} \in \mathbb{R}^3$ and equipped with $N_{\mathrm{BS}}$ antennas, a single-antenna user equipment (UE) at position $\mathbf{P}_{\mathrm{UE}} \in \mathbb{R}^3$, and a set $\mathcal{E}$ of $M = |\mathcal{E}|$ \glspl{EMS} (electromagnetic surfaces) placed at positions $\{\mathbf{P}_j^{\mathrm{EMS}}\}_{j=1}^M$, all expressed in a global reference system.
Each EMS consists of $L$ sub-wavelength meta-atoms located at $\{\mathbf{p}_{j,\ell}^{\mathrm{EMS}}\}_{\ell=1}^{L}$ relative to the center $\mathbf{P}_j^{\mathrm{EMS}}$. 

The detailed procedure for channel charting and position inference from \gls{csi} will be described in {Sec.~\ref{sec:cc_method}.

\subsection{Signal Model}
\label{subsec:signal_model}
\begin{figure}[t!]
    \centering
    \includegraphics[width=0.8\linewidth]{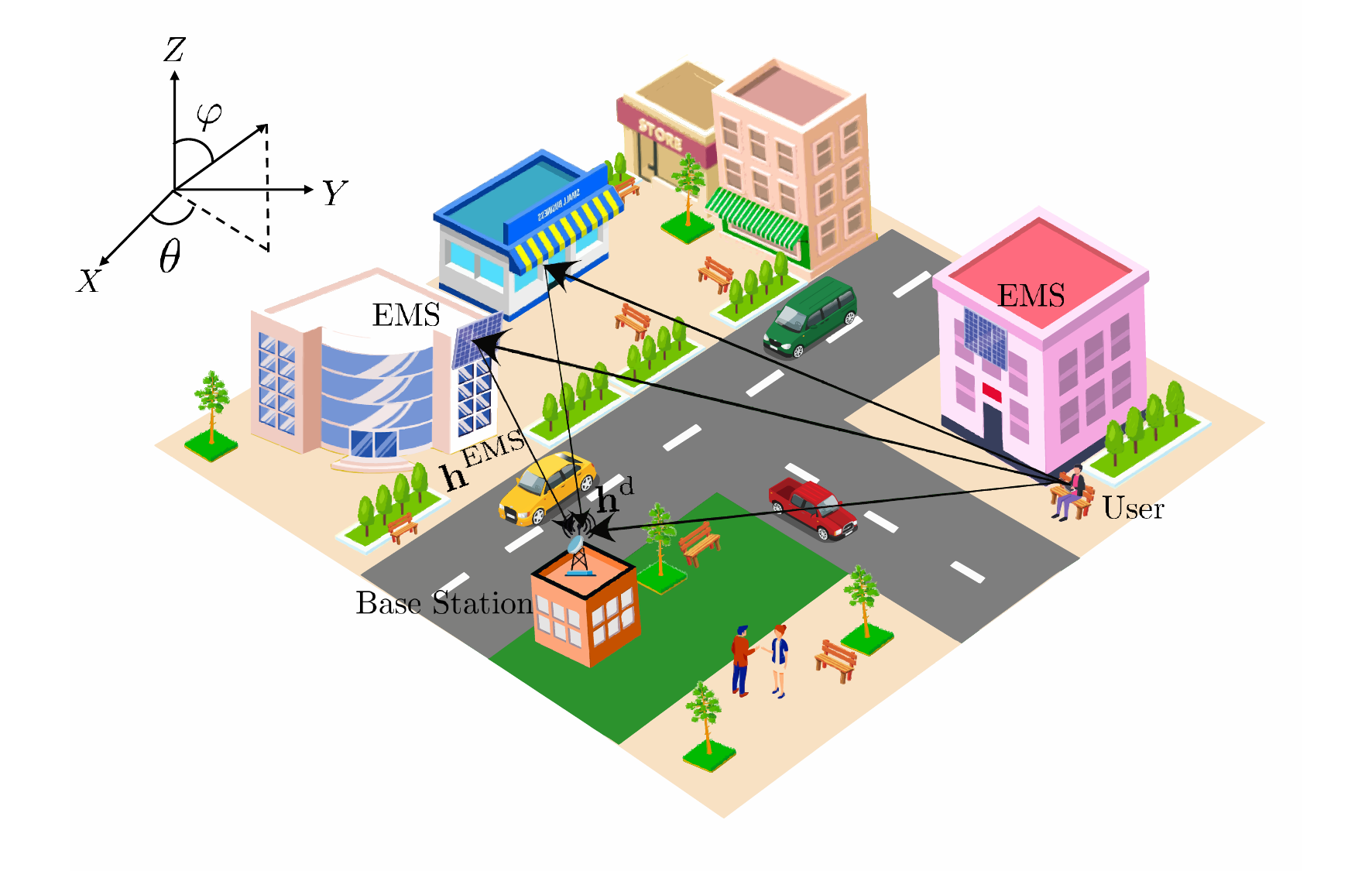}
  \caption{System model.}
  \label{fig:sysmodel}
\end{figure}

Let $s \in \mathbb{C}$ be the transmit symbol from the UE with $\mathbb{E}[|s|^2] = \sigma_s^2$. The time-discrete received signal at the BS is modeled as
\begin{equation}
    \mathbf{y}_{\mathrm{rx}} = \mathbf{h}(\mathcal{S}) \, s + \mathbf{n},
    \label{eq:rx_signal}
\end{equation}
where $\mathbf{y}_{\mathrm{rx}} \in \mathbb{C}^{N_{\mathrm{BS}} \times 1}$ is the received vector, $\mathbf{n} \sim \mathcal{CN}(\mathbf{0}, \sigma_n^2 \mathbf{I}_{N_{\mathrm{BS}}})$ is the complex Gaussian noise, and $\mathbf{h}(\mathcal{S}) \in \mathbb{C}^{N_{\mathrm{BS}} \times 1}$ is the composite channel vector, which depends on the set of EMS reflection configurations $\mathcal{S} = \{\boldsymbol{\Phi}_1, \dots, \boldsymbol{\Phi}_M\}$.

The channel vector is the superposition of the direct path and the reflected paths via all EMS:
\begin{equation}
    \mathbf{h}(\mathcal{S}) = \mathbf{h}^{\mathrm{d}} + \sum_{j \in \mathcal{E}} \mathbf{h}_j^{\mathrm{EMS}}(\boldsymbol{\Phi}_j),
    \label{eq:total_channel}
\end{equation}
where $\mathbf{h}^{\mathrm{d}} \in \mathbb{C}^{N_{\mathrm{BS}} \times 1}$ is the direct channel between UE and BS, and $\mathbf{h}_j^{\mathrm{EMS}}(\boldsymbol{\Phi}_j) \in \mathbb{C}^{N_{\mathrm{BS}} \times 1}$ is the contribution of the $j$-th EMS.
Each EMS is configured by a diagonal reflection matrix $\boldsymbol{\Phi}_j \in \mathbb{C}^{L \times L}$:
\begin{equation}
    \boldsymbol{\Phi}_j = \mathrm{diag}\left( e^{j\phi_{j,1}}, \ldots, e^{j\phi_{j,L}} \right),
    \label{eq:ems_phi}
\end{equation}
where $\phi_{j,\ell}$ is the phase shift introduced by the $\ell$-th element of the $j$-th EMS. This modeling is widely adopted in the literature and assumes negligible amplitude variation and inter-element coupling.
The channel contribution via the $j$-th EMS is given by
\begin{equation}
    \mathbf{h}_j^{\mathrm{EMS}}(\boldsymbol{\Phi}_j) = \mathbf{H}_j^{\mathrm{o}} \, \boldsymbol{\Phi}_j \, \mathbf{h}_j^{\mathrm{i}},
    \label{eq:ems_channel}
\end{equation}
where $\mathbf{h}_j^{\mathrm{i}} \in \mathbb{C}^{L \times 1}$ is the channel vector from the UE to the $L$ elements of EMS $j$, and $\mathbf{H}_j^{\mathrm{o}} \in \mathbb{C}^{N_{\mathrm{BS}} \times L}$ is the channel matrix from the $L$ elements of EMS $j$ to the $N_{\mathrm{BS}}$ antennas at the BS.

\subsection{Channel Model}
\label{subsec:channel_model}
We assume a block-fading channel with independent fading for each link. In this work, both the direct and EMS-assisted channels are modeled as deterministic multipath propagation, using physics-based ray tracing. Specifically, we employ the open-source Sionna Ray Tracing engine~\cite{SionnaRT}, which allows for detailed electromagnetic simulation in realistic environments.

The geometric scenario, including the positions and geometries of the BS, UE, EMSs, and relevant scatterers, is built using Blender, an open-source 3D modeling tool. This 3D environment is imported into Sionna RT~\cite{SionnaRT}, which simulates the propagation environment and outputs, for each link, a set of $P$ deterministic multipath components. Each path $p$ is characterized by a complex gain $\alpha_p$, departure and arrival angles $\boldsymbol{\vartheta}^p = (\theta^p, \varphi^p)$, path length, and delay.

The resulting channel impulse response is constructed as
\begin{equation}
    \mathbf{h} = \frac{1}{\sqrt{P}} \sum_{p=1}^{P} \alpha_p \, \varrho(\boldsymbol{\vartheta}^p) \, \mathbf{a}(\boldsymbol{\vartheta}^p),
    \label{eq:generic_channel}
\end{equation}
where $\varrho(\boldsymbol{\vartheta}^p)$ is the element radiation pattern, modeled as in~\cite{3GPP} for BS antennas and as in~\cite{CIRS} for EMS meta-atoms, and $\mathbf{a}(\boldsymbol{\vartheta}^p) \in \mathbb{C}^{N \times 1}$ is the array response vector, with $N = N_{\mathrm{BS}}$ for the BS and $N = L$ for the EMS.

The array response vector is given by
\begin{equation}
    \mathbf{a}(\boldsymbol{\vartheta}) = \left[e^{j \mathbf{k}(\boldsymbol{\vartheta})^\top \mathbf{p}_1}, \ldots, e^{j \mathbf{k}(\boldsymbol{\vartheta})^\top \mathbf{p}_N}\right]^\top,
    \label{eq:array_response}
\end{equation}
where $\mathbf{k}(\boldsymbol{\vartheta}) \in \mathbb{R}^{3 \times 1}$ is the wave vector,
\begin{equation} \label{eq:wavevector}
    \mathbf{k}(\boldsymbol{\vartheta}) = \frac{2\pi}{\lambda} 
    \begin{bmatrix}
        \cos(\varphi)\cos(\theta),
        \cos(\varphi)\sin(\theta),
        \sin(\varphi)
    \end{bmatrix},
\end{equation}
and $\mathbf{p}_n \in \mathbb{R}^{3}$ is the global position of the $n$-th antenna or meta-atom. All coordinates, including those for the BS, UE, and EMS elements, are referenced in the same global coordinate system for unambiguous modeling.

\section{Channel Charting Method}\label{sec:cc_method}

\gls{cc} aims to learn a low-dimensional representation of the spatial relationships between channel states, leveraging features derived from CSI. To enable both training and evaluation, we consider a set of \glspl{tp} $\mathcal{U} = \{\mathbf{p}_1, \ldots, \mathbf{p}_{N_\mathcal{U}}\}$, where each $\mathbf{p}_u \in \mathbb{R}^3$ denotes a possible UE location. Not all TPs correspond to active UEs at any given time, but CSI is collected for each.

\subsection{CSI Feature Construction and Dissimilarity Metrics}

A central step in CC is the extraction of features that capture the distinguishing spatial characteristics of the channel. In this work, we adopt the channel covariance matrix as the CSI feature. The covariance, a large-scale statistic, evolves slowly with position and can be estimated robustly in practice~\cite{kazemi2023beam}. For each test point $u$, we compute the covariance matrix as $\mathbf{R}_u(\mathcal{S}) = \mathbb{E}\left[\mathbf{h}_u(\mathcal{S}) \mathbf{h}_u^H(\mathcal{S})\right]$, where the expectation is taken over fading, multipath, and estimation errors (modeled as SNR-dependent noise~\cite{STMM}). Here, $\mathcal{S}$ denotes the EMS configuration. To quantify the dissimilarity between two channel states, we employ the Log-Euclidean (LE) distance between covariance matrices, which is effective for comparing high-dimensional Hermitian matrices:
\begin{align}
    d^{\mathrm{LE}}_{u,u'}(\mathcal{S}) &= \|\log \mathbf{R}_u(\mathcal{S}) - \log \mathbf{R}_{u'}(\mathcal{S})\|_F \\ \notag
    &= \sqrt{\operatorname{Tr}(\boldsymbol{\Lambda}(\mathcal{S}) \boldsymbol{\Lambda}^H(\mathcal{S}))},
\end{align}
where $\log(\cdot)$ denotes the matrix logarithm (computed via SVD~\cite{kazemi2023beam}), and $\boldsymbol{\Lambda}(\mathcal{S}) = \log \mathbf{R}_u(\mathcal{S}) - \log \mathbf{R}_{u'}(\mathcal{S})$. This dissimilarity metric forms the basis of the channel chart.

\subsection{Nonlinear Dimensionality Reduction: \textit{t}-SNE}

To embed the dissimilarity structure into a low-dimensional chart, we use \gls{tsne}~\cite{kazemi2023beam}. \gls{tsne} operates by matching the probability distributions of pairwise similarities in the high-dimensional feature space and the low-dimensional latent space.

Let $\mathbf{D}(\mathcal{S}) \in \mathbb{R}^{N_\mathcal{U}\times N_\mathcal{U}}$ be the matrix of LE distances. For each $u$, similarities to all other points are defined via a Gaussian kernel:
\begin{equation}
p_{u|u'}(\mathcal{S}) = \frac{\exp\left(-[\mathbf{D}(\mathcal{S})]_{u,u'}^2/2\sigma_u^2\right)}{\sum_{w \neq u} \exp\left(-[\mathbf{D}(\mathcal{S})]_{u,w}^2/2\sigma_u^2\right)},
\end{equation}
with $\sigma_u$ chosen such that the conditional probability distribution $p_{u|u'}$ achieves a specified \emph{perplexity}, a user-selected parameter that determines the effective number of nearest neighbors considered for each point and thus balances the preservation of local and global structure. 
%
%
%
%

In the low-dimensional latent space, we seek an embedding $\mathcal{Z} = \{\mathbf{z}_u\}_{u=1}^{N_\mathcal{U}} \subset \mathbb{R}^{d_{\text{lat}}}$, where each $\mathbf{z}_u$ is the image of test point $u$ in the latent space of dimension $d_{\text{lat}}$ (typically $d_{\text{lat}}=2$ or $3$). To quantify the similarity between pairs of latent points $(u, u')$, t-SNE employs a heavy-tailed Student-$t$ distribution (with one degree of freedom) centered at each point. By using the Student-$t$ kernel, \gls{tsne} effectively allocates more area in the latent space to represent moderate and large pairwise distances, thus preserving both local and some global data structure.

Specifically, the similarity between latent points $u$ and $u'$ is defined as:
\begin{equation}
    q_{u,u'} = \frac{\left(1 + \|\mathbf{z}_u - \mathbf{z}_{u'}\|^2\right)^{-1}}{\displaystyle\sum_{w \neq v} \left(1 + \|\mathbf{z}_w - \mathbf{z}_v\|^2\right)^{-1}},
\end{equation}
where the numerator assigns higher similarity to closer points, and the denominator normalizes the values over all distinct pairs $(w, v)$ in the dataset.

The objective of \gls{tsne} is to arrange the latent points $\{\mathbf{z}_u\}$ such that the distribution of pairwise similarities $Q = \{q_{u,u'}\}$ in the latent space matches as closely as possible the target similarity distribution $P = \{p_{u,u'}\}$ derived from the high-dimensional feature space. This is formalized as the minimization of the Kullback–Leibler (KL) divergence from $P$ to $Q$:
\begin{equation}
    \hat{\mathcal{Z}}(\mathcal{S}) = \arg\min_{\mathcal{Z}} \sum_{u,u'} p_{u,u'}(\mathcal{S}) \log \frac{p_{u,u'}(\mathcal{S})}{q_{u,u'}},
\end{equation}
\vspace{1mm}

where $p_{u,u'}(\mathcal{S})$ are the joint probabilities from the primary (feature) space, and $q_{u,u'}$ are those in the latent space.

This objective is optimized using gradient descent. The gradient of the KL divergence for a single latent coordinate $\mathbf{z}_u$ is given by:
\begin{equation}
    \frac{\partial f_{\text{t-SNE}}}{\partial\mathbf{z}_u} =
     4\sum_{u'} \left(p_{u,u'}(\mathcal{S}) - q_{u,u'}\right)
     \frac{\mathbf{z}_u - \mathbf{z}_{u'}}{1 + \|\mathbf{z}_u - \mathbf{z}_{u'}\|^2}.
\end{equation}
This gradient forces latent points to move closer together when their similarity in the primary space is underrepresented in the latent space, and to move apart when their similarity is overrepresented. Optimization proceeds by iteratively updating the latent coordinates $\mathcal{Z}$ to minimize the divergence.

\paragraph*{Semi-supervised t-SNE for Localization.}

Standard \gls{tsne} is unsupervised, meaning the latent coordinates $\{\mathbf{z}_u\}$ are determined solely by the structure of the input dissimilarity matrix and have no direct connection to real-world coordinates. To enable actual localization, we adopt a semi-supervised variant (St-SNE)~\cite{zhang2021semi}, in which the latent positions of a subset of labeled points $\mathcal{I} \subset \mathcal{U}$ are “clamped” to their known physical coordinates $\{\mathbf{y}_i\}_{i \in \mathcal{I}}$ throughout the optimization process. This acts as a set of anchor points, guiding the remaining (unlabeled) embeddings to align with the true spatial geometry, while still preserving the local and global structure imposed by the dissimilarities. 

During each iteration, only the unlabeled latent embeddings $\{\mathbf{z}_u : u \notin \mathcal{I}\}$ are updated via gradient descent, while the labeled points remain fixed. 
%
%

\section{Channel Charting in Smart Radio Environment}\label{sec:metrics_opt}
This section presents a framework for evaluating and optimizing \gls{cc} performance in the presence of \glspl{EMS}. We define key point-wise metrics, such as localization error, and mathematically parameterize EMS phase profiles to capture their impact on the \gls{csi} and resulting embeddings. The EMS configuration problem is then formulated as a non-convex, combinatorial optimization, for which we propose a tractable, codebook-based approach using a finite set of physically realizable phase patterns.

\subsection{Evaluation Metrics}\label{sec:EvaluationMetrics}

Let $\mathcal{U}$ denote the set of all test points (TPs), with $\mathcal{I} \subset \mathcal{U}$ representing labeled (anchor) points, and $\mathcal{L}' = \mathcal{U} \setminus \mathcal{I}$ the unlabeled points used for evaluation. For each $u \in \mathcal{L}'$, we define the following point-wise metrics:

\textbf{Localization Error (LE):}
\begin{equation}
    \mathrm{LE}_u(\mathcal{S}) = \big\| \hat{\mathbf{z}}_u(\mathcal{S}) - \mathbf{y}_u \big\|_2,
\end{equation}
where $\hat{\mathbf{z}}_u(\mathcal{S})$ is the latent embedding of point $u$ under EMS configuration $\mathcal{S}$, and $\mathbf{y}_u$ its true position.

\textbf{Trustworthiness (TW):}  
Let $\mathcal{V}_u(\kappa|\mathcal{S})$ and $\mathcal{V}'_u(\kappa|\mathcal{S})$ be the $\kappa$-nearest neighbors of $u$ in the primary and latent spaces, respectively. Then:
\begin{equation}
\mathrm{TW}_u(\kappa|\mathcal{S}) = 1 - \eta \sum_{\substack{u' \notin \mathcal{V}_u \\ u' \in \mathcal{V}'_u}} \big( r'_{u,u'} - \kappa \big),
\end{equation}
where $r'_{u,u'}$ is the rank of $u'$ in the latent neighbor list, and
\begin{equation}
    \eta = \frac{2}{\kappa(2|\mathcal{L}'| - 3\kappa - 1)}.
\end{equation}
TW $\in [0,1]$, with higher values indicating better local structure preservation.

\textbf{Continuity (CT):}  
\begin{equation}
\mathrm{CT}_u(\kappa|\mathcal{S}) = 1 - \eta \sum_{\substack{u' \in \mathcal{V}_u \\ u' \notin \mathcal{V}'_u}} \big( r_{u,u'} - \kappa \big),
\end{equation}
where $r_{u,u'}$ is the rank of $u'$ in the primary space. Like TW, CT $\in [0,1]$, with $1$ being ideal.

For any metric $m_u(\mathcal{S})$ (LE, $-\mathrm{TW}$, or $-\mathrm{CT}$ for minimization), we analyze its empirical distribution $F_m(x|\mathcal{S})$ and its $\alpha$-quantile $Q_m(\alpha|\mathcal{S})$. We adopt the strategy of \textit{optimizing the upper quantile} to improve worst-case performance, particularly in \gls{NLoS} regions.

\subsection{EMS Phase Profile Parameterization}

Let the incident and desired outgoing wave vectors be
\begin{equation}
  \mathbf{k}_i \triangleq \mathbf{k}(\boldsymbol{\vartheta}_i), \qquad
  \mathbf{k}_o \triangleq \mathbf{k}(\boldsymbol{\vartheta}_o),
\end{equation}
as defined in \eqref{eq:wavevector}. The generalized Snell’s law~\cite{CIRS} gives the required tangential phase gradient to achieve the desired reflection:
\begin{equation}
   \mathbf{k}_o - \mathbf{k}_i = \nabla_{\!\!\parallel} \Phi(\mathbf{r}) + \nu(\mathbf{r})\,\mathbf{u}(\mathbf{r}),
\end{equation}
with $\nabla_{\!\!\parallel}$ denoting the tangential gradient, $\Phi(\mathbf{r})$ the phase profile, and $\nu(\mathbf{r})$ a Lagrange multiplier for the normal component. For a planar EMS, this reduces to
\begin{equation}
   \Phi(\mathbf{r}) = \Phi_0 + (\mathbf{k}_o - \mathbf{k}_i)^\top \mathbf{r},
\end{equation}
which can be sampled on the discrete EMS elements as
\begin{equation}
   \phi_\ell = (\mathbf{k}_o - \mathbf{k}_i)^\top \mathbf{p}_\ell + \Phi_0.
\end{equation}

\subsection{Optimization Problem: Continuous and Codebook-Based Formulation}\label{OptimizationSection}

The goal is to find the EMS phase configuration $\mathcal{S}$ that minimizes the $\alpha$-quantile of the LE, negative TW, or negative CT evaluated over all test points. Explicitly,
\begin{equation}
    \hat{\mathcal{S}} = \arg\min_{\mathcal{S}\,\in\,\mathbb{S}}\;Q_{{m}}(\alpha|\mathcal{S}),
\end{equation}
where $\mathbb{S}$ is the set of all feasible phase matrices for all EMSs. For $M$ EMSs of $L$ elements each, $\mathbb{S}$ is the $M \times L$ dimensional torus of elementwise phase shifts:
\begin{equation}
    \mathbb{S} = \left\{ \{\Phi_j\}_{j=1}^M : \Phi_j = \diag(e^{j\boldsymbol{\phi}_j}),\; \boldsymbol{\phi}_j \in [0,2\pi)^L \right\}.
\end{equation}
This problem is high-dimensional, non-convex, and combinatorial, and thus intractable for practical EMS sizes. The objective function is highly non-linear in $\mathcal{S}$ due to the complex dependency of the channel and embedding on the EMS phase profile, and it have many local minima.

To make optimization tractable and align with EMS fabrication constraints, we adopt a \textit{codebook-based approach} \cite{Dai2020}.\vspace{1mm}
Here, a finite codebook $\mathcal{C}$ of $K$ candidate phase profiles is constructed—typically using a range of linear phase gradients or pre-selected angle pairs. The joint codebook for all $M$ EMSs is the Cartesian product $\mathbb{C} = \mathcal{C}^M$, with $K^M$ possible configurations:
\begin{equation}
    \hat{\mathcal{S}}
    = \arg\min_{\mathcal{S}\,\in\,\mathbb{C}} Q_{m}(\alpha|\mathcal{S}).
\end{equation}
This approach allows for \textit{exhaustive or greedy search} over a manageable set of profiles, enabling practical design. While it does not guarantee global optimality, it strikes a balance between computational tractability, hardware feasibility, and worst-case performance~\cite{Dai2020}.

\subsection{Codebook Construction for EMS Phase Profiles}

The codebook $\mathcal{C}$ is constructed to provide a finite set of physically realizable EMS phase profiles for tractable optimization and fabrication. In this work, we adopt a \emph{phase-gradient codebook} for EMS design, in which each candidate phase profile corresponds to a linear phase ramp across the EMS surface. Specifically, for a planar EMS with element positions $\mathbf{p}_\ell = (x_\ell, y_\ell)$ and grid spacings $d_x, d_y$, we define sets of quantized phase increments $\mathcal{C}_x = \{\Delta\phi_x^{(a)}\}_{a=1}^{K_x}$ and, $\mathcal{C}_y = \{\Delta\phi_y^{(b)}\}_{b=1}^{K_y}$. Each codeword is specified by a pair of discrete slopes $(a,b)$, and the corresponding phase profile is
\begin{equation}
    \phi_\ell^{(a,b)} = \Phi_0^{(a,b)} + \gamma_x^{(a)} x_\ell + \gamma_y^{(b)} y_\ell,
\end{equation}
where $\gamma_x^{(a)} = \Delta\phi_x^{(a)}/d_x$ and $\gamma_y^{(b)} = \Delta\phi_y^{(b)}/d_y$. In our experiments, we focus on 1D horizontal phase gradients ($\gamma_y^{(b)}=0$), so each codeword simplifies to
\begin{equation}
    \phi_\ell^{(a)} = \Phi_0^{(a)} + \gamma_x^{(a)} x_\ell.
\end{equation}
\begin{figure}[t!]
    \centering
    \includegraphics[width=0.7\linewidth]{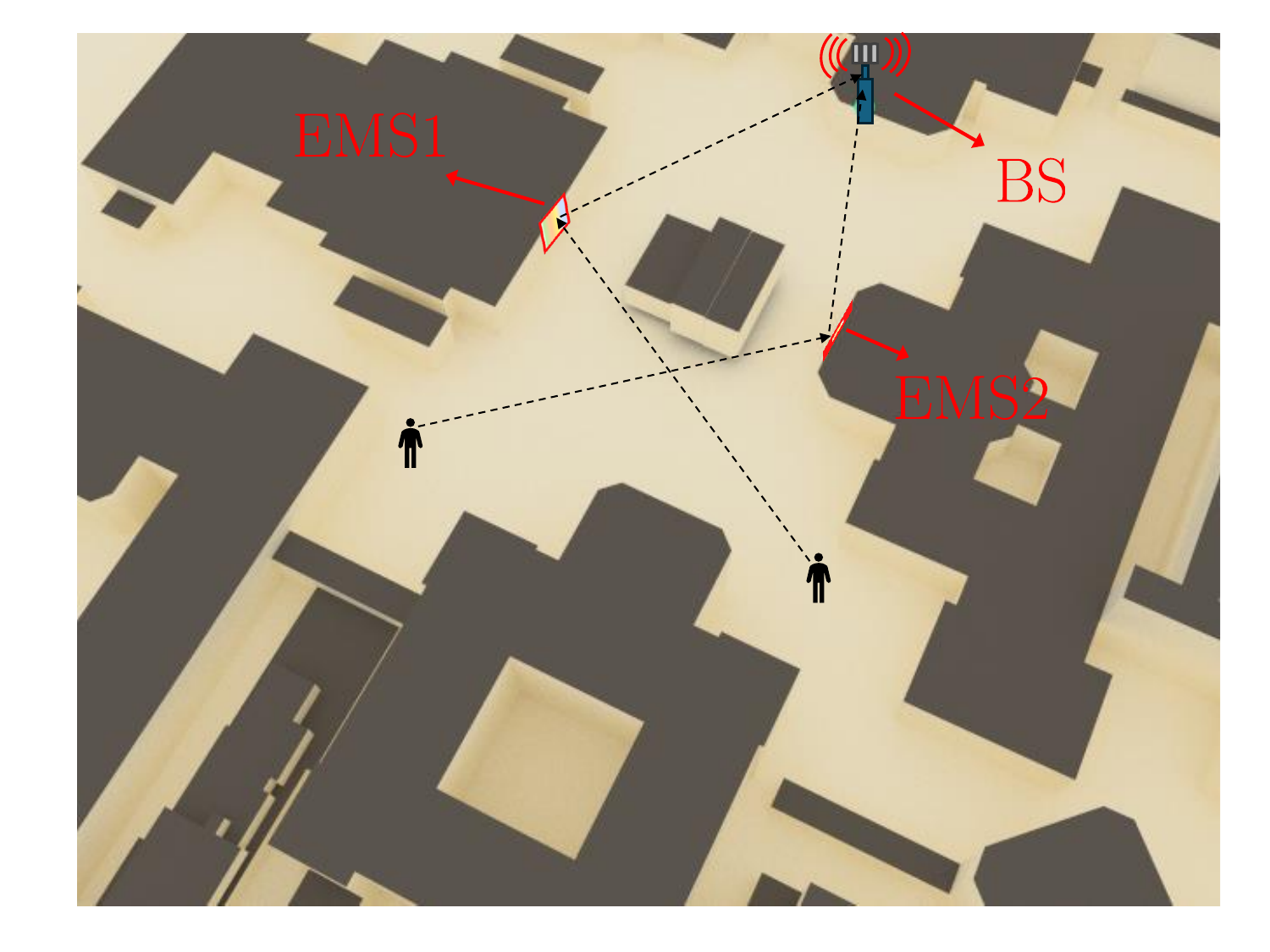}
  \caption{Geometry of the considered scenario.}
  \label{fig:topology}
\end{figure}
The resulting codebook $\mathcal{C}$ consists of $K=K_x$ unique phase profiles, each corresponding to a different steering direction or angular spread. For each EMS, the candidate phase matrices $\Phi^{(a)} = \diag(e^{j\phi_1^{(a)}}, \dots, e^{j\phi_L^{(a)}})$ are precomputed. For a system with $M$ EMSs, the joint search space is the Cartesian product $\mathbb{C} = \mathcal{C}^M$. In practice, the codebook parameters are selected to cover the range of anticipated propagation directions, with cardinality $K$ determined by the tradeoff between design complexity, hardware constraints, and optimization granularity~\cite{Dai2020}.

\section{Results and Discussion}
\label{sec:sim_results}

This section presents a comprehensive study of the proposed CC framework with EMSs, in a realistic urban deployment. We provide both quantitative and qualitative evaluations, highlight the influence of codebook-based EMS design, and discuss the performance trade-offs observed under various panel configurations.

The scenario, illustrated in Fig.~\ref{fig:topology}, spans an $80 \times 110\,\mathrm{m}^2$ urban area derived from \gls{OSM}. Channels are generated deterministically using the Sionna ray tracer, with path gains ($\alpha_p$) and physical multipath properties accurately reflected in the input CSI features. The \gls{BS} is placed on top of the tallest building, while two static $60 \times 60$ EMSs are mounted on facing building walls at $5.5$~m height. A total of 3200 TPs are uniformly distributed to represent potential user locations, covering both \gls{LoS} and \gls{NLoS} regions. All simulations are done with 15\% supervision ratio. Key simulation parameters are summarized in Table~\ref{tab:var}.

\begin{table}[b!]
    \centering
    \footnotesize
    \caption{Default simulation parameters.}
    \begin{tabular}{l|c|c}
    \toprule
        \textbf{Parameter} &  \textbf{Symbol} & \textbf{Value(s)}\\
        \hline
        Carrier frequency & $f_0$  & $30\,\mathrm{GHz}$ \\
        Bandwidth & $B$ & $10\,\mathrm{MHz}$\\
        UE transmit power & $\sigma_s^2$ & $23\,\mathrm{dBm}$\\
        Noise power & $\sigma_n^2$ & $-92\,\mathrm{dBm}$\\
        EMS size & $L \times L$ & $60 \times 60$\\
        EMS element spacing & $d_n,d_m$ & $\lambda_0/4$ \\
        BS antenna array & $N_\mathrm{BS}$ & $8\times 4$\\
        UE antenna array & $N_\mathrm{t}$ & $1$ \\
        Tx/Rx element spacing & $d_\mathrm{Tx},d_\mathrm{Rx}$ & $\lambda_0/2$ \\
        BS height & $h_\mathrm{BS}$ & $8.5\,\mathrm{m}$ \\
        UE height & $h_\mathrm{UE}$ & $1.5\,\mathrm{m}$ \\
        EMS height & $h_\mathrm{EMS}$ & $5.5\,\mathrm{m}$ \\
        \bottomrule
    \end{tabular}
    \label{tab:var}
\end{table}

For each \gls{EMS}, a codebook of 11 DFT-based horizontal phase gradients is used. These codewords are empirically designed to ensure good coverage of \gls{NLoS} regions, and the DFT structure guarantees slope orthogonality given the EMS size. Horizontal orientation is chosen for simplicity and computational efficiency. While this is a simplification, it is justified by the optimization scale and aligns with practical EMS fabrication constraints.

A total of $11 \times 11 = 121$ two-panel codeword combinations are evaluated. Increasing the codebook size yields no significant improvement, confirming the adequacy of the current discretization. Further optimization—such as joint placement and codebook design—requires network planning and is left for future work.

\begin{figure}[thb!]
  \centering
  \includegraphics[width=0.8\linewidth]{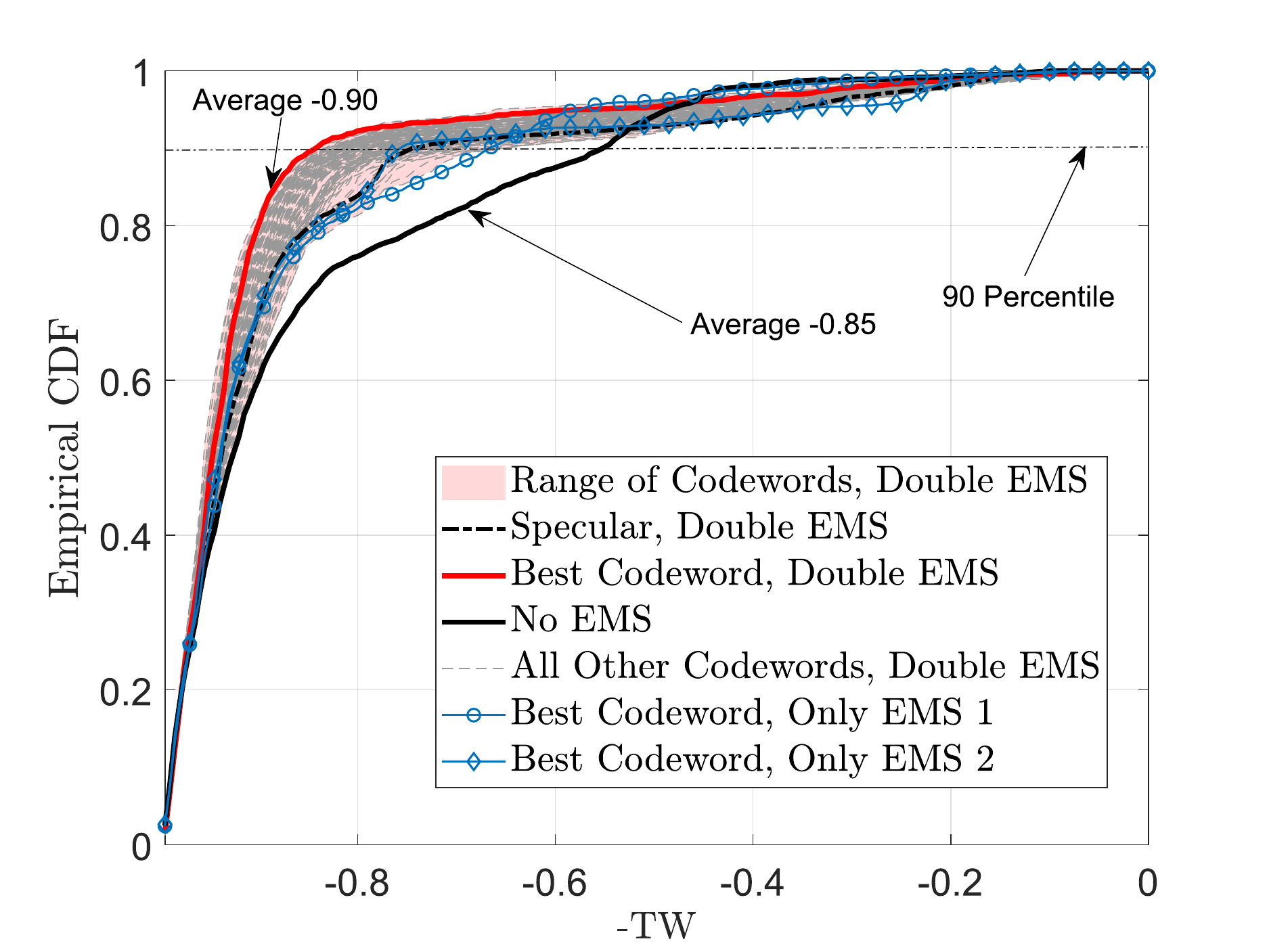}
  \caption{Empirical CDF of $-\mathrm{TW}$.}
  \label{fig:TW}
\end{figure}

\begin{figure}[thb!]
  \centering
  \includegraphics[width=0.8\linewidth]{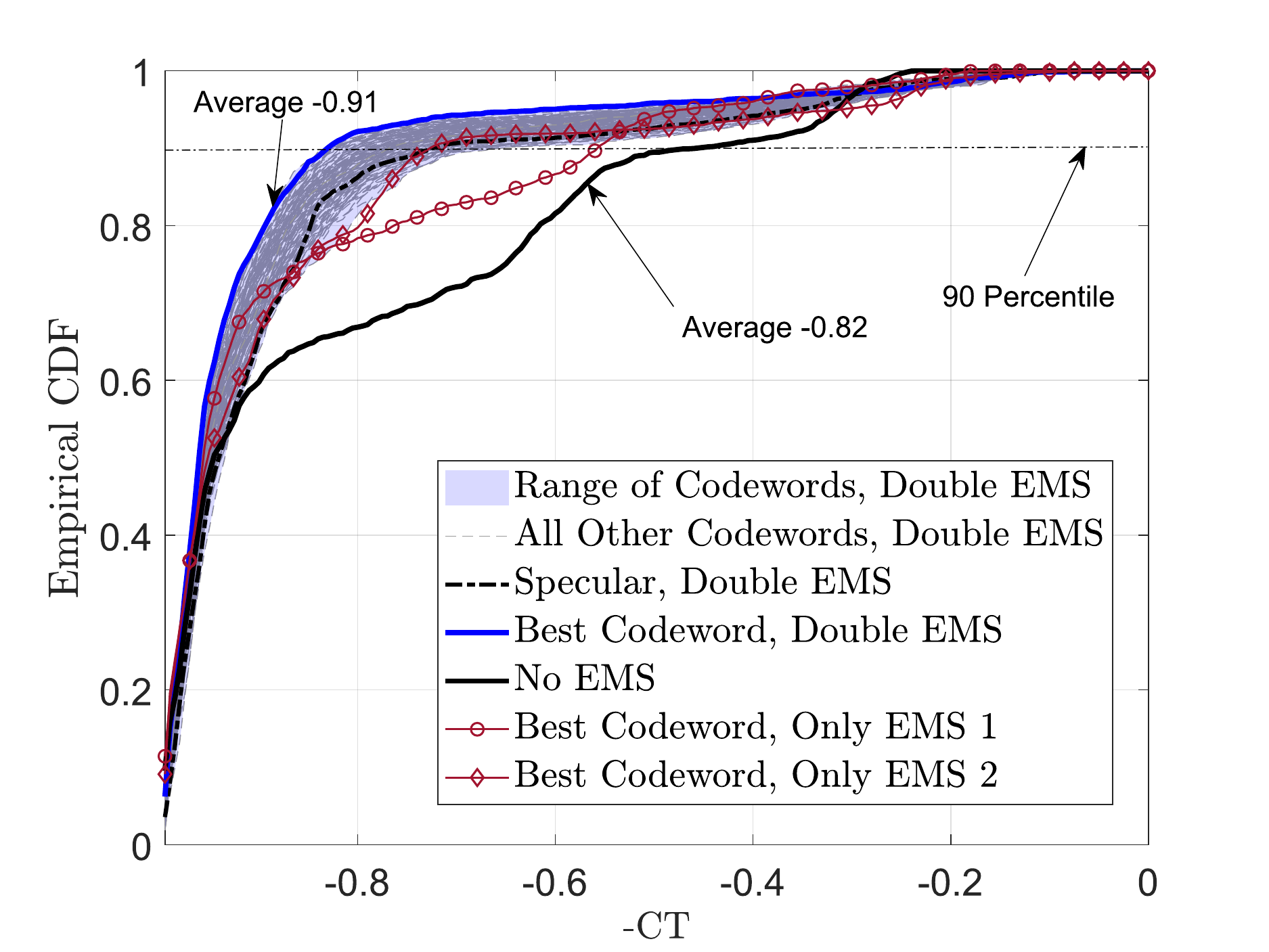}
  \caption{Empirical CDF of $-\mathrm{CT}$}
  \label{fig:CT}
\end{figure}

\begin{figure}[thb!]
  \centering
  \includegraphics[width=0.8\linewidth]{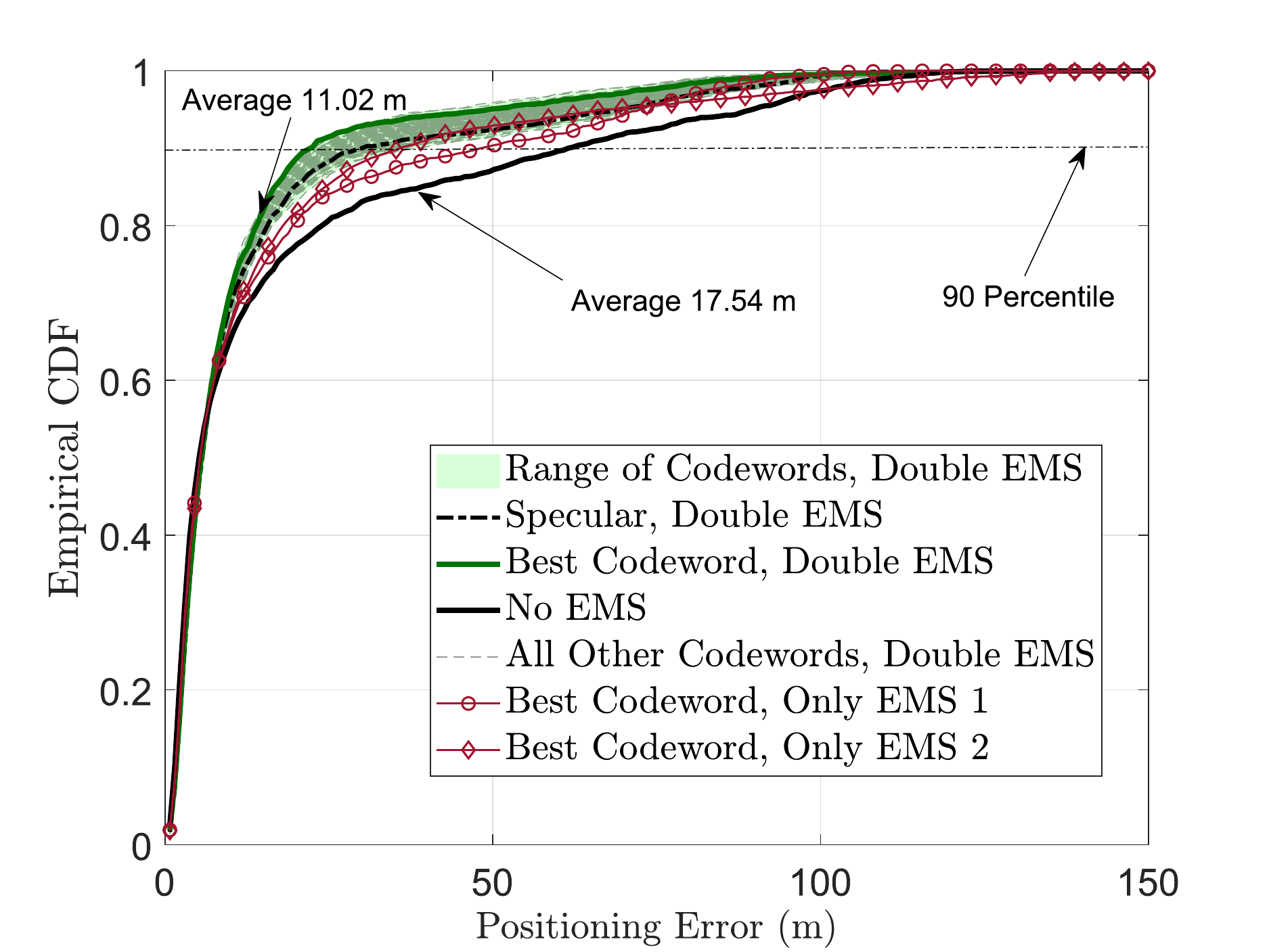}
  \caption{Empirical CDF of positioning error.}
  \label{fig:PosError}
\end{figure}

Figures~\ref{fig:TW}--\ref{fig:PosError} show empirical \glspl{CDF} of trustworthiness, continuity, and positioning error across all test points. 
The compared configurations include: (i) the baseline with no \gls{EMS} (solid black), (ii) specular \glspl{EMS} acting as mirrors (dash-dotted), (iii) the best codebook configuration optimizing the 90th percentile (solid color), (iv) the full envelope across all 121 codeword combinations (gray band), and (v) the best single-panel case, where only \gls{EMS} 1 or \gls{EMS} 2 is active (markers).

\begin{figure}[th!]
  \centering
  \includegraphics[width=.48\linewidth]{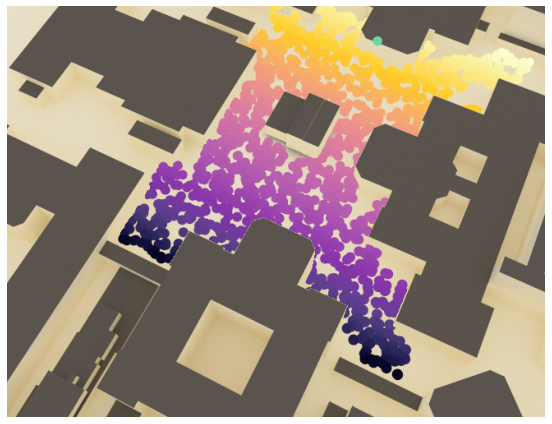}
  \includegraphics[width=.49\linewidth]{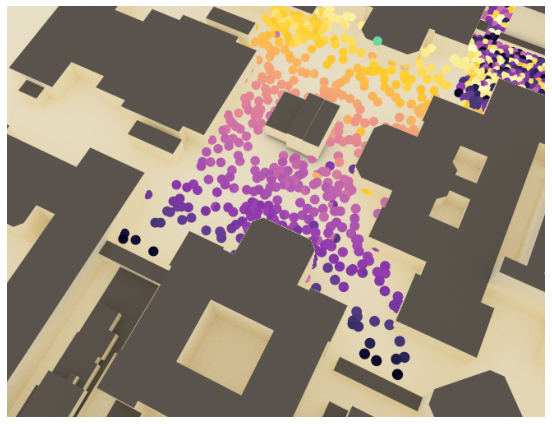}\\
  \includegraphics[width=.49\linewidth]{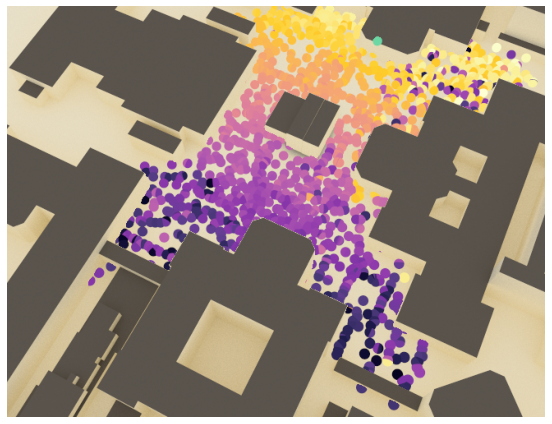}
  \includegraphics[width=.49\linewidth]{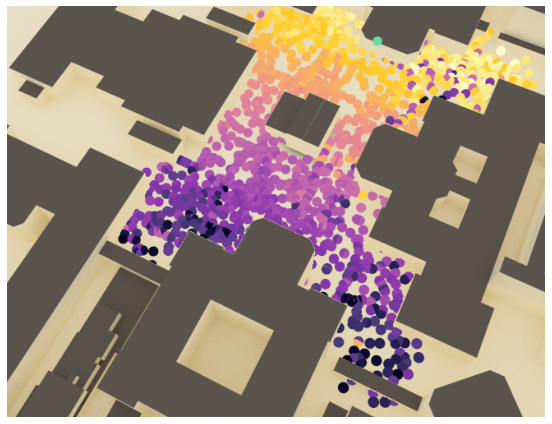}
  \caption{Channel chart embeddings: (a) ground truth positions, (b) no EMS, (c) specular EMS, (d) best codebook EMSs. Colors reflect $y$-coordinates. Only codebook-optimized EMSs recover full spatial geometry, especially in NLoS.}
  \label{fig:colorbar}
\end{figure}
Among all pre-configuration themes, codebook-optimized EMSs yield the greatest gains in the 60th–95th percentiles—corresponding to \gls{NLoS} or difficult user positions. For the lowest-error users, improvements are limited due to already favorable direct paths. No configuration performs worse than the baseline, confirming that EMS deployment consistently improves performance, particularly in challenging cases. These improvements are most pronounced for hard-to-localize users and are sensitive to EMS placement and codebook selection. While minor for mostly \gls{LoS} users, the gains in \gls{NLoS} regions underscore the potential of static \glspl{EMS}. With the best codeword, the 90-th percentile of localization error can be decrease from above 60 meters, to less than 25 meters, only with the help of fully passive \gls{EMS}s. Achieving optimal performance, however, would require joint network planning and co-design of EMS placement and codebooks, which is left for future work.

Figure~\ref{fig:colorbar} provides a qualitative comparison of the embedding quality. The true $y$-coordinate is color-coded, allowing a direct visual check of how well spatial relationships are preserved.

Without \glspl{EMS}, the \gls{NLoS} regions collapse into tight clusters in the embedding, destroying the spatial ordering. Specular (mirror-like) panels partially recover the structure, but local errors remain—especially near building edges. Only the best codebook-optimized \glspl{EMS} enable a faithful mapping: color bands are uniform and well ordered, with strong separation of distant points even in challenging \gls{NLoS} zones.

\section{Conclusion}
This paper presented a scenario-driven framework for enhancing channel charting (CC) through the integration of static electromagnetic surfaces (EMS) in smart radio environments. By enriching the multipath structure of wireless channels, EMS significantly improve localization accuracy, trustworthiness, and continuity—especially in non-line-of-sight (NLoS) conditions. We formulated EMS phase profile design as a codebook-based optimization problem targeting upper quantiles of key embedding metrics. This approach is both tractable and physically realizable. 3D ray-traced simulations in a realistic urban setting confirmed that optimized EMS configurations can significantly decrease the taio of the localization error and substantially improve embedding quality over specular and baseline cases. These results confirm that static, pre-configured \glspl{EMS}, when carefully optimized, provide robust gains for challenging user locations without requiring real-time reconfiguration.

\bibliographystyle{IEEEtran}
\bibliography{IEEEabrv,biblio}

\end{document}